\documentclass[lettersize,journal]{IEEEtran}
\usepackage{amsmath,amsfonts}
\usepackage{algorithmic}
\usepackage{algorithm}
\usepackage{array}
\usepackage[caption=false,font=normalsize,labelfont=sf,textfont=sf]{subfig}
\usepackage{textcomp}
\usepackage{stfloats}
\usepackage{url}
\usepackage{verbatim}
\usepackage{graphicx}
\usepackage{cite}
\usepackage{booktabs}
\usepackage{amssymb}

\begin{document}

\title{Understanding the power-law nature of participation in community sports organizations}

\author{Jia~Yu, Mengjun~Ding, Weiqiang~Sun, ~\IEEEmembership{Senior~Member,~IEEE}, \\ Weisheng Hu, ~\IEEEmembership{Member,~IEEE}, Huiru~Wang
\thanks{Manuscript received June, 2023. (Corresponding author: Weiqiang Sun.)}
\thanks{Jia Yu, Mengjun Ding, Weiqiang Sun, and Weisheng Hu are with the School of Electronic Information and Electrical Engineering, Shanghai Jiao Tong University, Shanghai 200240, China. Huiru Wang is with the Department of Physical Education, Shanghai Jiaotong University, Shanghai 200240, China.
(e-mail: \{yujia543, mengjun\_ding, sunwq, wshu, wanghr\}@sjtu.edu.cn).}
}



\maketitle

\begin{abstract}
The improvement of living standards and awareness of chronic diseases have increased the importance of community sports organizations in promoting the physical activity levels of the public. However, limited understanding of human behavior in this context often leads to suboptimal resource utilization. In this study, we analyzed the participation behavior of 2,956 members with a time span of 6 years in a community sports organization. Our study reveals that, at the population level, the participation frequency in activities adheres to a power-law distribution. To understand the underlying mechanisms driving crowd participation, we introduce a novel behavioral model called HFBI (Habit-Formation and Behavioral Inertia), demonstrating a robust fit to the observed power-law distribution. The habit formation mechanism indicates that individuals who are more engaged are more likely to maintain participation, while the behavioral inertia mechanism suggests that individuals' willingness to participate in activities diminishes with their absences from activities. At the individual level, our analysis reveals a burst-quiet participation pattern, with bursts often commencing with incentive activities. We also find a power-law distribution in the intervals between individual participations. Our research offers valuable insights into the complex dynamics of human participation in community sports activity and provides a theoretical foundation to inform intervention design. Furthermore, the flexibility of our model enables its application to other data exhibiting power-law properties, broadening its potential impact beyond the realm of community sports.
\end{abstract}


\section{Introduction}
\IEEEPARstart{G}{lobalization} urbanization, and increased wealth have led to significant lifestyle changes, causing a wide decrease in physical activity. According to the World Health Organization (WHO), inactivity rates can climb as high as 70\% in certain countries, primarily due to shifts in transportation habits, heightened reliance on technology, and urbanization \cite{world2019global}. Physical inactivity, which has been identified as a global pandemic, is responsible for up to 8\% of non-communicable diseases and deaths globally \cite{katzmarzyk2022physical,kohl2012pandemic}. Conservatively estimated, physical inactivity cost health-care systems INT\$53.8 billion worldwide in 2013 \cite{ding2016economic}. Additionally, if the prevalence of physical inactivity remains unchanged, it is projected that by 2030, there will be around 499.2 million new cases of preventable major NCDs worldwide, resulting in direct health-care costs of INT\$ 520 billion. The annual global cost of not taking action on physical inactivity is anticipated to reach approximately \$47.6 billion \cite{santos2023cost}. 

In an effort to improve physical activity participation, community sports organizations have achieved remarkable results in recent years. Many concur that community sport, as a low-threshold physical activity, is a powerful tool for targeting socially vulnerable groups \cite{van2020community}. Moreover, community sport has been recognized as a policy area and a social field that goes beyond ``just" providing opportunities for groups to participate in sports. It also encompasses functions such as social care and crime reduction \cite{debognies2019personal,schaillee2019community}. Today, being non-profit by nature, community sports organizations face greater challenges, such as competition for limited resources, volunteer availability, and capacity, and the impact of pandemics (such as COVID-19) \cite{doherty2022return}. Understanding the nature of the population participating in community sports is thus pivotal to making the best use of limited resources.

The interest in the data-driven exploration of human behavior has been persistent. Very early on, power-law distribution has been found in certain human behaviors, such as the intervals between emails \cite{barabasi2005origin}, the pattern of phone calls \cite{jiang2013calling,jo2012circadian,jiang2016two}, and complex social networks \cite{broido2019scale,eom2011characterizing,rak2020fractional,anderson2014maximizing}. Efforts have been made to understand the principle behind the formation of this power-law distribution in these behaviors \cite{newman2005power,kumamoto2018power}. Classical models such as the decision-based queuing process \cite{barabasi2005origin} and preferential attachment \cite{barabasi1999emergence} are proposed to explain the power law distribution observed in the waiting time for processing emails and the degree distribution in complex networks, respectively. Research on community sports organizations is usually conducted from an organizational management perspective, providing high-level guidance for organizational development by quantifying aspects such as resources, program design, diversity, life cycle, and resilience \cite{doherty2019organizational,anderson2014maximizing,spaaij2018diversity}. However, very few, if any, models are population-based and consider when, how, and who participates in community-level sports activities \cite{westerbeek2021physical}. 

In this study, with the data from 2,956 users collected over a span of six years, we discovered a power-law distribution of population participation in community sports activities. To explain this power-low distribution, we proposed the hypothesis of habit formation and behavioral inertia in community sports activity participation. Previous research has indicated that physical activity behavior can be developed through repeated experience of the activity in stable contexts \cite{hagger2019habit,lally2010habits}. Human behavior does exhibit inertia, as evidenced by the tendency for users to stick with default options \cite{liu2017limit} and purchase habits \cite{henderson2021customer}. Our empirical data provides evidence of habit formation and behavioral inertia in community sports participation. It may help to address the question, ``What is the typical `shape' of within-person real-world habit growth with repetition over the long-term" identified in the 2019 European Health Psychology Society Synergy Expert Meeting \cite{gardner2021developing}. Based on these two mechanisms we designed a behavioral model called HFBI that can robustly fit the power-law distribution of the empirical data. Power-law distribution is also observed in the interval of participation at the individual level, signifying a burst-quiet pattern of activity participation. With the relevant activity information, we found that bursts tend to be initiated by activities with incentive rewards, suggesting that incentive activities can help call people back for sustained engagements. The main contributions of the article as described as follows.
\begin{enumerate}
\item{For the first time, we discovered that the frequency of population participation in community physical activities and the interval between individual's participations obey power-law distributions.}

\item{We proposed an intuitive model to explain the power-law distribution of population participation in community physical activities, by taking into account habit formation and behavioral inertia. We demonstrated good fitting performance and statistical significance with real-world data. The model may as well be used in other domains where power-law distributions with low power-law exponents are observed.} 

\item{The intervals between individual's participation exhibit a power-law distribution, with a pattern of bursts followed by periods of inactivity (a burst-quiet pattern). We observed that bursts often start with incentive activities located in the head position. This implies that incentive activities not only attract more participants but also have the potential to call users back from a quiet state to an active state, thereby promoting sustained engagement.}
\end{enumerate}

The rest of this article is organized as follows. In Section II, we demonstrate the power-law phenomenon of participation frequency in activities at the population level. In Section III, we introduce the proposed HFBI model and present the evidence. In Section IV, we verify the participation patterns at the individual level and the role of incentive activities. In Section V, we present the related work. Finally, we summarize this paper in Section VI.

\begin{figure*}[!t]
\centering
\subfloat[]{\includegraphics[width=0.4\linewidth]{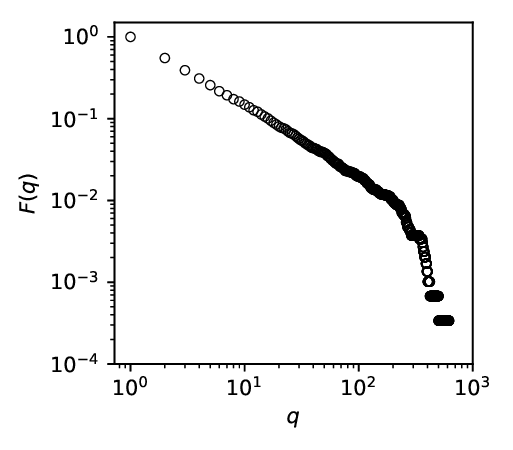}%
\label{fig:figure1_A}}
\hfil
\subfloat[]{\includegraphics[width=0.4\linewidth]{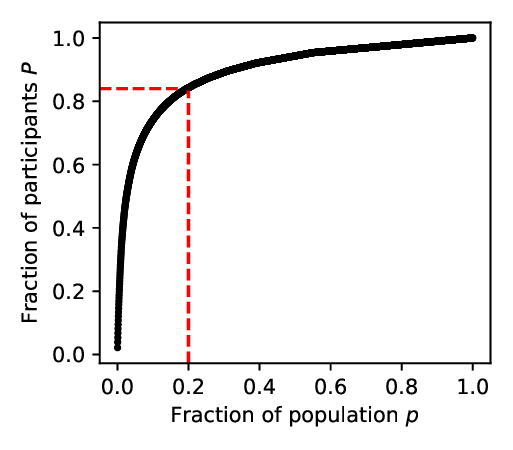}%
\label{fig:figure1_B}}
\caption{Power law of participation frequency at the population level. (a) Complementary cumulative probability distribution of the participation frequency. $q$ represents the frequency of participation, and $F(q)$ denotes the proportion of members who participate in activities with a frequency not less than $q$. (b) The fraction $P$ of the total participation in the community sports organization held by the fraction $p$ of the most active members. It can be seen that the most active 20\% of the population hold about 84\% of the participations (red dashed lines).}
\label{fig:figure1}
\end{figure*}

\begin{figure}[!b]
\centering
\includegraphics[width=.8\linewidth]{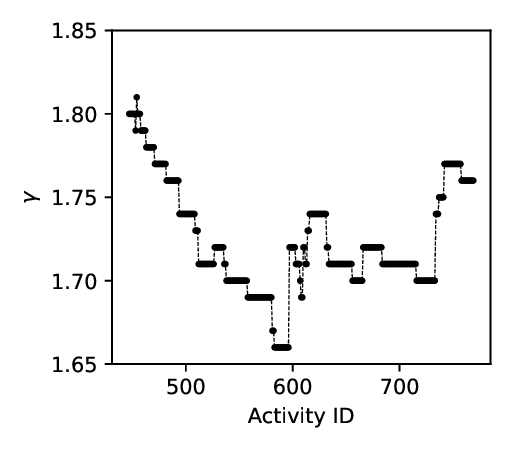}
\caption{$\gamma$ of the power-law distribution of the participation frequency at each activity node after the platform reached 1000 users (Activity ID 447). The $p$-values of KS tests are all greater than 0.1.}
\label{fig:figure2}
\end{figure}

\section{Power-law Distribution of Participation Frequency at the Population Level}
\subsection{Data Description} The data used in our research was sourced from a university-based community sports platform that we develop and operate, which allows individuals to initiate or participate in sports activities. The initiator of the activity can choose whether or not to provide rewards as incentives for the activity. Over the course of 6 years, from May 2015 to May 2021, our dataset captured 28,714 records of activity participation in 770 activities (including 110 activities with incentives), involving a total of 2,956 individuals. Each record in the dataset contains the participant's ID, activity ID, team ID, and type of activity (whether to provide incentives or not). The activity IDs are consecutive natural numbers starting from 0 and arranged in the order of their occurrence (numbered from 0 to 769).

\subsection{Fitting the Empirical Data}
The frequency of user $i$ participating in activities over the entire period is denoted as $q_i$. For the sequence of activity participation frequency $\left\{ q_{i} \right\}$, we assume that the frequency larger than a truncated value $q_{min}$ is described by the power law distribution, 

\begin{equation}
p(q) \sim q^{-\gamma}, q \geq q_{\min }.
     \label{} 
\end{equation}

In the Kolmogorov-Smirnov (KS) test, $p>0.1$ (or $p>0.05$) suggests that the data can be considered to follow a power-law distribution. We select the smallest value of $q$ that satisfies the KS test with $p>0.1$ as $q_{min}$, and the data above $q_{min}$ can be plausibly modeled as a power-law distribution. The estimate $\gamma$ is chosen by maximum likelihood (MLE) \cite{broido2019scale,clauset2009power}. 


\subsection{Power-law Distribution of Participation Frequency}
The participation frequency of the population follows a power-law distribution. Fig. \ref{fig:figure1_A} shows the empirical distribution of user participation frequency in activities in a complementary cumulative way to enhance the statistical significance \cite{barabasi2013network}. The complementary cumulative function can be represented as $F(q)=\sum_{q^{\prime}=q}^{\infty} p({q^{\prime})}$, where $p(q)$ denotes the proportion of individuals who participated in activities $q$ times. A clear straight-line trend can be observed on the double logarithmic axis, indicating a power law distribution of the data. Kolmogorov–Smirnov (KS) tests and Maximum likelihood estimation (MLE) fits are employed to check whether the empirical distributions obey power law distribution and estimate the related parameters. The result shows that the frequency of population participation in the activity is in line with the power law distribution ($p=0.18$, $q_{min}=2$) with the power-law exponent $\gamma=1.76$. The cutoff of the tail indicates that there are fewer individuals participating in an exceptionally large number activity than what a power-law distribution would expect, which is a phenomenon commonly observed in real-world systems. Fig. \ref{fig:figure1_B} shows the relationship between the fraction $P$ of the participation and the most active $p$ of the population. 80/20 rule is evident that the top 20\% of the most active users contributed to approximately 84\% of the total activity participation. Theoretically, the case is more extreme for power-law distributions with $\gamma$ less than 2. However, the fact that the number of activities is finite and the tail cutoff brings the ratio close to the classical Pareto's law.

To demonstrate that the power-law distribution of the participation frequency is not momentary coincidental, we analyzed the data for each activity node after the platform scale reached 1000. All samples (287 (88.9\%) with $q_{min}$=1 and 36 (11.1\%) with $q_{min}$=2) conformed to the power law distribution by KS test, with $p$-values all greater than 0.1. Fig. \ref{fig:figure2} presents the $\gamma$ for all samples of 323 activity nodes. The range of $\gamma$ spans from 1.66 to 1.81 with a mean of 1.72. And it keeps changing slowly with each activity held, first decreasing steadily, and then fluctuating and rising. The $\gamma$ less than 2 indicates a significant ``heavy tail" phenomenon in the frequency of participation.

\begin{figure*}[!t]
\centering
\subfloat[]{\includegraphics[width=0.4\linewidth]{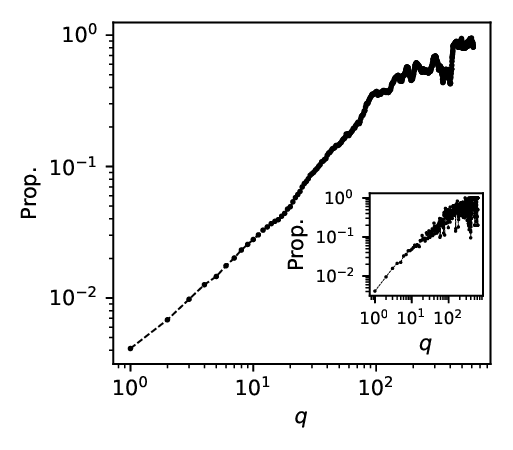}%
\label{fig:figure3_A}}
\hfil
\subfloat[]{\includegraphics[width=0.4\linewidth]{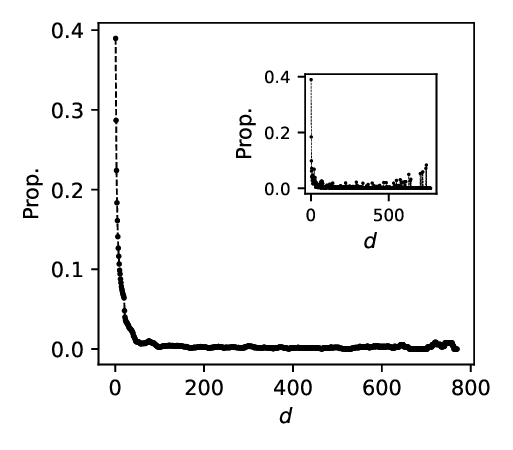}%
\label{fig:figure3_B}}
\caption{Evidence of habit formation and behavioral inertia. (a) The proportion of people who choose to continue participating in an available activity varies according to the number of activities they have participated in. $q$ denotes the number of activities that have been participated in before the currently available activity. (b) The proportion of people who choose to continue participating in an available activity varies depending on the number of activities they have been absent from. $d$ represents the number of sessions between the last activity they participated in and the currently available one. To better visualize the trend, a smoothing process with a 20-point window is applied to the results, and the original results are presented in the small inset chart. There are some small occasional fluctuations due to small samples when $q$ and $d$ are large.}
\label{fig:figure3}
\end{figure*}

\section{HFBI-A Behavioral Model Based on Habit Formation and Behavioral Inertia}
To explore the principle behind the power-law distribution of the participation frequency, we propose a behavioral model named HFBI, which is based on the assumptions of habit formation and behavioral inertia. Intuitively, people who have participated in activities frequently or have just participated in an activity are more likely to participate in subsequent activities. They are supported by convincing evidence from our data. 

\subsection{Evidence for Habit Formation and Behavioral Inertia}To provide evidence for the habit formation and behavioral inertia mechanisms, we performed a statistical analysis of all activities in the dataset. The proportion of people who have participated in $q$ activities and would choose to participate in a new available activity can be represented as

\begin{equation}
\operatorname{prop} .(q)=\frac{\sum_{j=0}^{N-1} m_{q}^{j}}{\sum_{j=0}^{N-1} n_{q}^{j}}.
     \label{} 
\end{equation}
Here, $n_{q}^{j}$ represents the number of individuals who have participated in $q$ activities before a new activity $j$, $m_{q}^{j}$ represents the number of individuals among them who choose to participate in the activity $j$, and $N$ is the total number of activities in the dataset. The denominator represents the total number of individuals who have participated in $q$ activities for all activities, while the numerator represents the number of individuals who choose to continue to participate in an activity after participating in $q$ activities. 

Similarly, the proportion of people who have been away from activities for $d$ sessions and would choose to participate in a new available activity can be represented as 

\begin{equation}
\operatorname{prop} .(d)=\frac{\sum_{j=0}^{N-1} m_{d}^{j}}{\sum_{j=0}^{N-1} n_{d}^{j}}.
    \label{} 
\end{equation}
$n_{d}^{j}$ represents the number of individuals who have been away from activities for $d$ sessions for activity $j$, $m_{d}^{j}$ represents the number of individuals among them who choose to participate in the activity $j$. The denominator represents the total number of individuals who have been away from activities for $d$ sessions for all activities, while the numerator represents the number of individuals who choose to continue to participate in an activity after being away from activities for $d$ sessions.

Fig. \ref{fig:figure3_A} shows the proportion of people who have participated in $q$ activities and would choose to participate in a new available activity. As shown, the proportion of individuals opting to continue participation increases almost linearly with the number of activities participated in the early stage. Fig. \ref{fig:figure3_B} illustrates the proportion of people who have been away from activities for $d$ sessions and would choose to participate in a new available activity. As the number of sessions away from activities increases, the proportion of people choosing to back to participating in activities sharply decreases. These provide solid evidence for the existence of habit formation and behavioral inertia in community sports participation.

\subsection{The HFBI Model}
Based on the evidence presented, we propose the HFBI model, which incorporates habit formation and behavioral inertia, to simulate user participation in activities. The experimental results demonstrate that the model can accurately simulate user participation in activities with only four parameters.
\subsubsection{Parameter Settings}The HFBI model only requires four parameters: $n$, $c$, $m$,  and $\alpha$. $n$ represents the number of activities held, i.e., the model's iteration count. $c$ and $m$ refer to the quantities of new and existing users participating in an activity (added in a round of iterations), respectively. $\alpha$ is a parameter that adjusts the ratio of habit formation and behavioral inertia to achieve a better fit with the empirical data. The parameters of $c$ and $m$ can be derived from the mean values of the dataset. Note that since the parameters are natural integers, the values of $c$ and $m$ will be rounded. To ensure consistency in the scale of the population, $n$ is calculated based on the number of population, $c$, and $m$. Additionally, we initiate the iteration process with $m$ pre-existing users to enable the selection of existing users at the start of the iteration.

\subsubsection{Model Description}
The model is characterized by adding users in a sequential and batched manner, which aligns with many real-life situations. Initially, we make the assumption that for every activity, there will be $c$ new users and $m$ existing users participating. For a new available activity and an existing user $i$, $q_{i}$ represents the total number of activities that user $i$ has participated in before, and $d_{i}$ represents the interval between the last activity they participated in and the current new activity. User $i$ participating in the activity can be attributed to two mechanisms. (1) User $i$ has a probability of $\alpha$ to participate in the activity due to habit formation, which means the probability of participating is proportional to $q_{i}$:

\begin{equation}
\frac{q_{i}}{\sum_{i \in I} q_{i}}.
        \label{eqn:habit_formation} 
\end{equation}
(2) Additionally, there is a probability of $1-\alpha$ for user $i$ to participate in the activity due to behavioral inertia, which means the probability is a decreasing function of $d_{i}$: 

\begin{equation}
\frac{1 / d_{i}}{\sum_{i \in I} 1 / d_{i}}.
         \label{eqn:2} 
\end{equation}
Therefore, the total probability of user $i$ participating in the activity is: 

\begin{equation}
\phi_{i}=\alpha \frac{q_{i}}{\sum_{i \in I} q_{i}}+(1-\alpha) \frac{1 / d_{i}}{\sum_{i \in I} 1 / d_{i}}.
        \label{eqn:total_prop} 
\end{equation}
$I$ is the set of all existing users. The model will perform $n$ rounds of iterations, adding $c$ new users and selecting $m$ existing users based on Eq. \ref{eqn:total_prop} in each round. The $c$ new users will be added to the existing user pool in each round. The overall process of the model is shown in Algorithm \ref{alg:alg1}. Note that the specific form of the decreasing function for $d_i$ is not unique, as it can be adjusted by the parameter $\alpha$.

\begin{algorithm}[H]
\caption{The HFBI Model}\label{alg:alg1}
\begin{algorithmic}
\STATE 
\STATE {\textbf{Input:}} The number of new($n$) and existing($m$) users participating in each activity, the number of activities organized ($n$), and the model control parameter $\alpha$
\STATE {\textbf{Output:}} The sequence of participation frequencies for users $\left\{ q_{i} \right\}$
\STATE {\textbf{Initialization:}} existing users $I = \left\{0,1,2 ... m-1\right\}$, \\current active users $C = \varnothing$, $l_{i}=0$, $q_{i} = 0$ \textbf{ for } $i$ in $I$.
\STATE {\textbf{for}} $j=0$; $j<n$; $j++$ {\textbf{do}}
\STATE \hspace{0.5cm}$d_{i}=j-l_{i} $ \textbf{ for } $i$ in $I$
\STATE \hspace{0.5cm}$ \phi_{i}=\alpha \frac{q_{i}}{\sum_{i \in I} q_{i}}+(1-\alpha) \frac{1 / d_{i}}{\sum_{i \in I} 1 / d_{i}}$ \textbf{ for } $i$ in $I$
\STATE \hspace{0.5cm}$C \gets$ $c$ new users and $m$ users selected from $I$ based on $ \phi_{i}$
\STATE \hspace{0.5cm}$l_{i}=j$, $q_{i}++$ \textbf{ for } $i$ in $C$
\STATE \hspace{0.5cm}add $c$ new users to $I$
\STATE {\textbf{end for}}
\STATE \textbf{return} $\left\{ q_{i} \right\}$

\end{algorithmic}
\label{alg1}
\end{algorithm}

\subsubsection{Proof of Power-Law Distribution and Exponent in Habit Formation}
When only considering the habit formation, that is $\phi_{i}=\frac{q_{i}}{\sum_{i \in I} q_{i}}$, the model can generate power-law distribution data with a power exponent $\gamma=2+\frac{c}{m}$. The proof process is similar to the Price model \cite{newman2018networks}. In the HFBI, for every activity held, there will be $c$ new users and $m$ existing users participating, and the participation probability of existing users is proportional to the number of activities they have participated in before. Let $p_{q}(n)$ be the fraction of users that have participated $q$ times when the platform contains $n$ users, which is also the probability distribution of participation frequency. $q_i$ represents the number of activities participated by user $i$. When organizing an activity where only one user among all existing users will participate, the probability of existing user $i$ participating in the activity is

\begin{equation}
\frac{q_{i}}{\sum_{i} q_{i}}=\frac{q_{i}}{n\langle q\rangle}=\frac{q_{i}}{n \frac{m+c}{c}}.
    \label{} 
\end{equation}
where $\langle q\rangle$ represents the average number of activities each person participates in, $\langle q\rangle=n^{-1} \sum_{i} q_{i}$. The number of people who have participated in $q$ activities is $np_{q}(n)$. When there is a new activity, the expected number of people who have participated in $q$ activities and will join the new activity is

\begin{equation}
n p_{q}(n) \times m \times \frac{q}{n \frac{m+c}{c}}=p_{q}(n) \times m \times \frac{q}{\frac{m+c}{c}}.
       \label{} 
\end{equation}

Then the master equation for the evolution of the participation frequency distribution is 

\begin{equation}
(n+c) p_{q}(n+c)=n p_{q}(n)+\frac{(q-1) m c}{m+c} p_{q-1}(n)-\frac{m q c}{m+c} p_{q}(n).
     \label{eqn:master} 
\end{equation}

The left side of the equation is the expected number of people participating in the activity $q$ times after adding an activity. The first term on the right-hand side here represents the number of users with previous $q$ participation. The second term refers to the expected number of users who have a participation frequency of $q-1$ and join the activity and become $q$ times, while the third term refers to the expected number of users who have a participation frequency of $q$ and participate in this activity and are no longer $q$ times.

Eq. \ref{eqn:master} is applicable for all cases where $q \neq 1$. When $q = 1$, the right side of the equation will increase by $c$ new users whose participation frequency becomes 1, instead of the second term in Eq. \ref{eqn:master}, and the equation for $q=1$ is

\begin{equation}
(n+c)p_{1}(n+c)=n p_{1}(n)+c-\frac{m c}{m+c}p_{1}(n).
     \label{eqn:master1} 
\end{equation}

When considering the limit of large population size $n \rightarrow \infty$ and calculating the asymptotic form of the distribution participation frequency in this limit, we take the limit $n \rightarrow \infty$ and use the shorthand $p_{q}=p_{q}(\infty)$. Eqs. \ref{eqn:master} and \ref{eqn:master1} become 

\begin{equation}
p_{q}=\frac{(q-1) m c}{c(m+c)+m q c} p_{q-1} \quad \text { for } q>1,
      \label{eqn:5} 
\end{equation}

\begin{equation}
p_{1}=\frac{m+c}{2 m+c} \quad \text { for } q = 1.
      \label{} 
\end{equation}
Let $k = c/m$, then

\begin{equation}
p_{1}=\frac{1+k}{2+k} \quad \text { for } q = 1,
       \label{eqn:13} 
\end{equation}

\begin{equation}
p_{q}=\frac{(q-1)}{1+k+p} p_{q-1} \quad \text { for } q>1.
    \label{eqn:14}
\end{equation}

With Eqs. \ref{eqn:13} and \ref{eqn:14}, we can iteratively determine $p_{q}$ for all values of $q$, beginning with our initial solution for $p_{1}$. The results are as follows:

\begin{equation}
\begin{array}{l}
p_{1}=\frac{1+k}{2+k} \vspace{2ex}  \\
p_{2}=\frac{1}{2+k+1} \times \frac{1+k}{2+k}  \vspace{2ex} \\
p_{3}=\frac{2}{3+k+1} \times \frac{1}{2+k+1} \times \frac{1+k}{2+k} \vspace{2ex}\\
p_{4}=\frac{3}{4+k+1} \times \frac{2}{3+k+1} \times \frac{1}{2+k+1} \times \frac{1+k}{2+k} \vspace{2ex}\\
...\\
\end{array}
\end{equation}

The expression for general $q$ can be successively derived as:

\begin{equation}
p_{q}=\frac{(q-1) \times(q-2)  \ldots \times 1 \times(1+k)}{(q+k+1) \times(q-1+k+1) \ldots \times(2+k+1) \times(2+k)}.
       \label{eqn:16}
\end{equation}
It is known that the gamma function is

\begin{equation}
\Gamma(x)=\int_{0}^{\infty} t^{x-1} \mathrm{e}^{-t} \mathrm{~d} t,
\end{equation}
and it has the property that

\begin{equation}
\Gamma(x+1)=x \Gamma(x) \quad \text { for } x > 0.
\end{equation}
Applying this equation iteratively, we find that

\begin{equation}
\frac{\Gamma(x+n)}{\Gamma(x)}=(x+n-1)(x+n-2) \ldots x.
\end{equation}
Using this result, we can rewrite Eq. \ref{eqn:16} as 

\begin{equation}
p_{q}=(1+k) \frac{\Gamma(q) \Gamma(2+k)}{\Gamma(1) \Gamma(2+k+q)}.
 \label{eqn:20}
\end{equation}
By further employing Euler's formula 

\begin{equation}
B(x, y)=\frac{\Gamma(x) \Gamma(y)}{\Gamma(x+y)},
\end{equation}
Eq. \ref{eqn:20} can be simplified to 

\begin{equation}
p_{q}=\frac{(1+k)}{\Gamma(1)} B(q, 2+k) .
 \label{eqn:22}
\end{equation}

\begin{figure}[t]
\centering
\includegraphics[width=.8\linewidth]{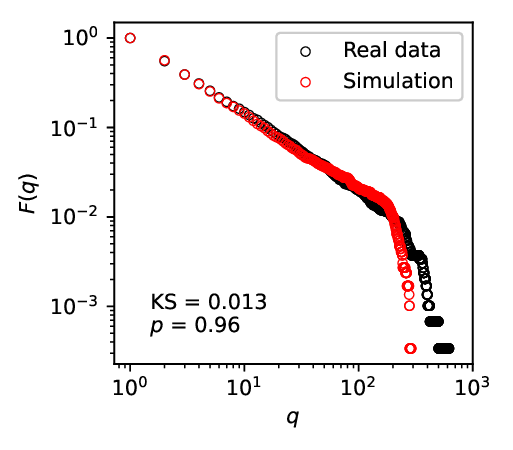}
\caption{Comparison of model-generated data and real data. The goodness-of-fit of the model is assessed by the KS tests. $p>0.1$ suggests that the two samples are likely to have originated from the same distribution.}
\label{fig:figure4}
\end{figure} 

\begin{figure*}[b]
\centering
\subfloat[]{\includegraphics[width=.4\linewidth]{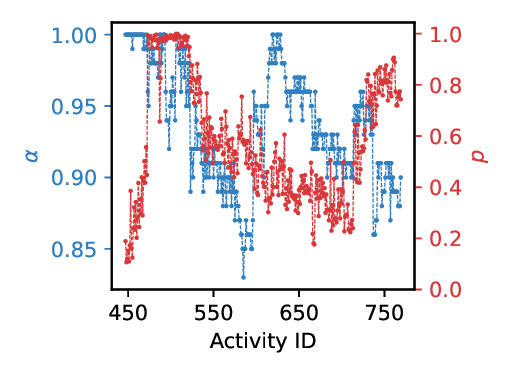}%
\label{fig:figure5_A}}
\hfil
\subfloat[]{\includegraphics[width=.4\linewidth]{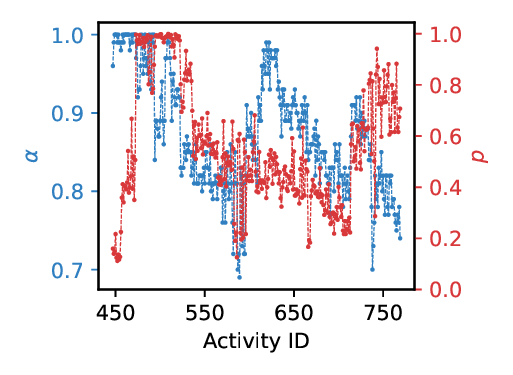}%
\label{fig:figure5_B}}
\caption{Figures (a) and (b) illustrate the optimal $\alpha$ of model fitting and average $p$-values with 5 runs
for two different equation representations of behavioral inertia: $\frac{1 / d_{i}}{\sum_{i \in I} 1 / d_{i}}$,$\frac{e^{-d_{i}}}{\sum_{i \in I} e^{-d_{i}}}$. $\alpha$ and $p$-value correspond to the blue and red y-axis, respectively.}
\label{fig:figure5}
\end{figure*}

Using Stirling’s approximation for the gamma
function, the beta function $B(x, y)$ falls off as a power law for large values of $x$, with exponent $y$ \cite{newman2018networks},

\begin{equation}
\mathrm{B}(x, y) \simeq x^{-y} \Gamma(y).
\end{equation}
Applying this finding to Eq. \ref{eqn:22}, for large values of $q$, the distribution of participation frequency goes as 

\begin{equation}
p_{q} \sim q^{-\gamma}=q^{-(2+k)}=q^{-(2+\frac{c}{m})},\\
        \label{} 
\end{equation}
where the exponent $\gamma$ is

\begin{equation}
\gamma=2+k=2+\frac{c}{m}. \\
     \label{} 
\end{equation}

Therefore, by only considering habit formation, represented by $\phi_{i}=\frac{q_{i}}{\sum_{i \in I} q_{i}}$, the model is able to generate data with a power-law distribution, where the power exponent is given by $\gamma=2+\frac{c}{m}$.

\subsubsection{Experimental Results on the Real Dataset}
We conducted experiments on real data, and the results show that HFBI is capable of generating data with only four parameters derived from the mean values of the empirical data and also exhibits good statistical significance. 

The Kolmogorov-Smirnov (KS) test is used to assess whether the data generated by the model and empirical data are drawn from the same distribution. The KS statistic is a value that measures the maximum distance between two cumulative distribution functions (CDFs) of two samples, which is used to determine if two samples are drawn from the same underlying probability distribution or not. The null hypothesis is that the two distributions are identical. If $p > 0.1$, we cannot reject the null hypothesis, which suggests that the data generating process is plausible. 

The experiment is first performed on the largest-scale data, that is, the data up to the last activity node. The parameter values for $c$, $m$, and $n$ are derived from the mean values of the data and are determined as 4, 33, and 731, respectively. In Fig. \ref{fig:figure4}, a comparison is shown between the generated data from HFBI and the real data. It can be seen that the distribution of the simulated data and the real data are very close. The model achieves the best fit when $\alpha$ is set to 0.9. The $\alpha$ values within the range of 0 to 1 suggest that the results of the empirical distribution are attributed to the combined effects of both habit formation and behavioral inertia mechanisms. The habit formation mechanism described by Eq. \ref{eqn:habit_formation} can be demonstrated to generate data with a power-law distribution for $\gamma=2+\frac{c}{m}$, which is strictly greater than 2 and differs from the empirical data. The participation frequency with $\gamma$ less than 2 implies that the frequency of participation in activities is slightly more than what can be explained by the habit formation mechanism alone. The behavioral inertia mechanism precisely compensates for this deficiency, as it captures the situation of individuals who have just participated in an activity being highly likely to continue participating in one or two due to inertia. It effectively adjusts the exponent while preserving the power-law distribution. It is the joint effect of both mechanisms that generate data that closely fit the empirical data. 

The data produced by the model is incapable of including the extremely rare users who have engaged in activities excessively. One possible explanation is that these individuals usually have a strong self-motivation to participate in activities, which cannot be captured by habit formation, as evidenced by the non-steady growth in the later stage of Fig. \ref{fig:figure3_A}. And since the parameters have to be integers and the operation to maintain consistency of the number of users between the generated data and the empirical data, there will be a small difference between the model's $n$ and the actual number of activity counts. This is considered acceptable since the proportion of these individuals is extremely low.


\begin{figure*}[!b]
\centering
\subfloat[]{\includegraphics[width=0.8\linewidth]{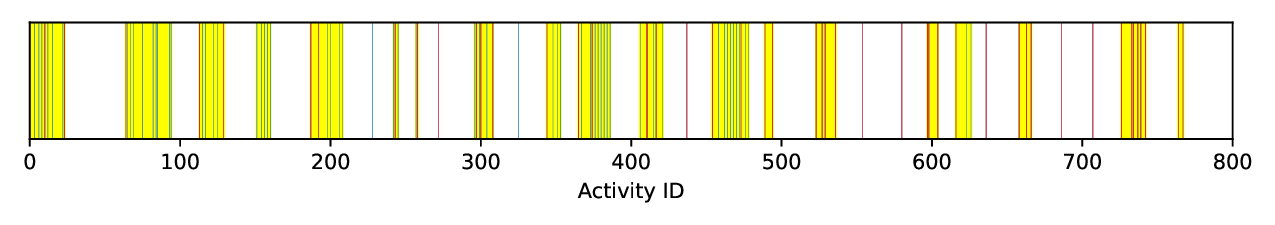}%
\label{fig:figure6_A}}
\hfil
\subfloat[]{\includegraphics[width=0.4\linewidth]{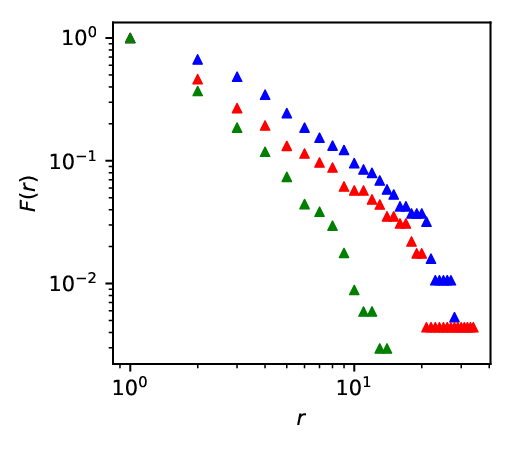}%
\label{fig:figure6_B}}
\hfil
\subfloat[]{\includegraphics[width=0.4\linewidth]{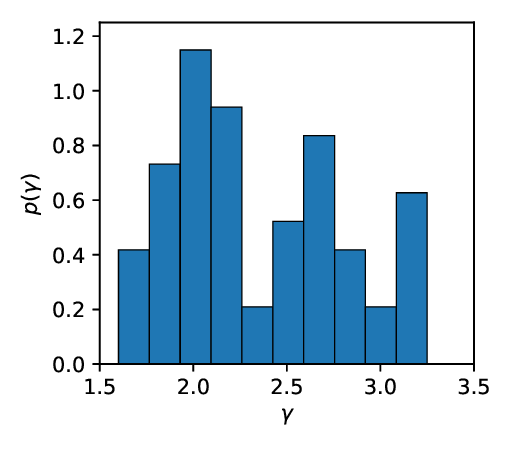}%
\label{fig:figure6_C}}
\caption{Patterns and characteristics of individual participation in activities. (a) An example of a real user's participation in activities. Vertical lines indicate the user's participation in the corresponding activity with the activity ID. The blue line represents ordinary activities, whereas the red line represents incentive activities. The yellow rectangular background represents an identified burst. (b) Complementary cumulative probability distributions of activity participation intervals for three real anonymous users. Different colors represent different users. (c) Probability distribution of estimated $\gamma$ for power-law distribution of activity participation intervals for 58 loyal users.}
\label{fig:figur6}
\end{figure*}

To demonstrate the robustness of the model, the model was also employed to fit the participation frequency up to each activity node. As the generated data can be slightly different each time, we conducted 5 runs for each possible value of $\alpha$ and selected the optimal $\alpha$ value with the highest average $p$-value among 5 runs. The average $p$-values and corresponding optimal $\alpha$ of model fitting for 323 samples are shown in Fig. \ref{fig:figure5}. In Fig. \ref{fig:figure5_A} and Fig. \ref{fig:figure5_B}, the behavioral inertia mechanism is represented by $\frac{1 / d_{i}}{\sum_{i \in I} 1 / d_{i}}$,$\frac{e^{-d_{i}}}{\sum_{i \in I} e^{-d_{i}}}$, respectively. It shows that different functional forms can achieve a good fit at different values of $\alpha$. The model shows good fitting performance ($p>0.1$) for all empirical data samples, indicating its correctness and robustness. The range of $\alpha$ values from 0.69 to 1 suggests that the proportion of habit formation and behavioral inertia may vary in different situations. We can observe clear downward trends in $\alpha$ around 450 to 600, indicating that the proportion of behavioral inertia gradually increases during this stage. By combining with Fig. \ref{fig:figure2}, it can be observed that there is also a decreasing trend of $\gamma$. This indicates that behavioral inertia can effectively help to capture situations with smaller $\gamma$.
\section{Participation Patterns at the Individual Level}

At the population level, the frequency of participation in activities follows a power-law distribution. At the individual level, the pattern of activity participation, specifically the intervals between each user's participation, is also worth studying. Similarly, we investigated the distribution of intervals between each individual's activity participation and discovered that they also exhibit a power-law distribution. In terms of activity participation patterns, it is a burst-quiet mode where individuals alternate between periods of high activity and periods of low activity. 

\subsection{The Burst-Quiet Pattern}
The interval between an individual's participation is defined as the subtraction of the IDs of two consecutive activities in which they have participated, denoted by the $r$. Considering the requirement of a sufficient amount of interval sequence data, we focused on 58 loyal users who participated in more than 100 activities for the individual-level analysis. Fig. \ref{fig:figure6_A} shows an example of a real user's participation in activities. It is evident that intervals of individual participation in activities vary greatly in size, with a majority being small and some being large. The participation of individuals is characterized by alternating bursts of high activity and long periods of low activity, similar to the outgoing mobile phone call sequence of an individual \cite{karsai2012universal}. This burst-quiet pattern is common among the group of loyal users. We studied the distribution of interval sequences for all 58 users and discovered that their interval sequences also follow a power-law distribution ($p>0.1$ for 54 users, $p>0.05$ for all 58 users, $r_{min}$=1 for 48 users, and $r_{min}$=2 for 10 users). 

The power law distribution also plays an important role in the intervals of individual participation in activities. Fig. \ref{fig:figure6_B} shows examples of complementary cumulative probability distributions of the intervals for three users. The intervals of participation in the activities of each of the three individuals obeyed a power law distribution with different power exponents. Fig. \ref{fig:figure6_C} plots the probability distribution of the estimated power-law exponents $\gamma$ for all loyal individuals, revealing a range from 1.6 to 3.25 and a mean of 2.35. Although their activity participation intervals all follow power-law distributions, the difference in the power-law exponent is quite significant. The range of $\gamma$ is surprisingly consistent with $\gamma$ for individuals with the intraday inter-call duration that follows a power-law distribution reported by Jiang \emph{et al} \cite{jiang2013calling}. And the probability distributions are also somewhat similar, which may suggest a potential connection between the intervals of different human behaviors.

\subsection{The Role of Incentive Activities in Bursts} 

Burst, characterized by frequent participation in activities with short intervals within a specific period, has a significant impact on improving individuals' overall fitness level. Therefore, it is important to explore the factors associated with this pattern to promote physical activity among the population. In this study, a burst is defined as a period in which the interval between consecutive activities a user participates in is less than a threshold value $\Delta$. The specific value of $\Delta$ is arbitrarily set in empirical analysis \cite{jiang2013calling}.

\begin{figure}[t]
\centering
\includegraphics[width=.8\linewidth]{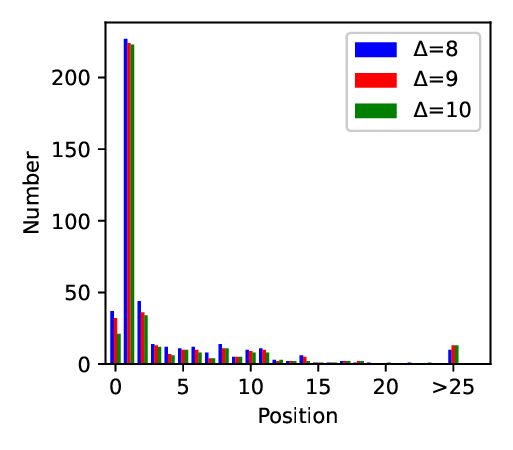}
\caption{The number of bursts with the first appearance of incentive activities at different positions within the burst. Position=0 indicates that no incentive activities are appearing in the burst. Blue, red, and green correspond to bursts identified under different $\Delta$ thresholds.}
\label{fig:figure7}
\end{figure}

Organizations often invest resources to provide incentives for activities to attract users to participate. Incentives are crucial in promoting physical activity. Typically, physical activity behavior is initially motivated by incentive, and as habits form, it shifts towards unconscious and automatic processes \cite{hagger2019habit}. The effectiveness of incentives can be immediately reflected in the number of participants in the activity. However, the benefits in other aspects are yet to be discovered. Our study has made some findings by observing the position of incentive activities in bursts. At thresholds of $\Delta$=8, 9, and 10, we identified a total of 433, 399, and 378 bursts for all individuals, respectively, and recorded the positions of the first occurrence of the incentive activity within each burst. As shown in Fig. \ref{fig:figure7}, the majority of bursts are observed to start with incentive activities. Table \ref{tab:tab} shows the number and percentage of bursts with the first incentive activity appearing at the head position in the bursts. Over 50\% of bursts have their first incentive activity in the first position, and over 65\% in the first three positions at different $\Delta$. Note that there is only one in seven activities is incentivized. The proportion of incentive activities in the head of bursts is much higher than it, indicating a correlation between the occurrence of incentive activities and bursts. This phenomenon suggests that in addition to increasing the number of participants in the activity, incentive activities may also play a role in calling users back from a quiet state to a burst state for sustained engagements.

\begin{table}[t]
\begin{center}
\centering
\caption{Bursts with First Incentive Activity at Different Positions \label{tab:tab}}
\begin{tabular}{lrrr}
\toprule
Type of Bursts & $\Delta=8$ & $\Delta=9$ & $\Delta=10$ \\
\midrule
Bursts with FIA at 1st pos.
 & 227(52.4\%)
 & 224(56.1\%)
 & 223(59.0\%)
 \\
Bursts with FIA at 1st 2 pos.
 & 271(62.6\%)
 & 260(65.2\%)
 & 257(68.0\%)
 \\
Bursts with FIA at 1st 3 pos.
 & 285(65.8\%)
 & 273(68.4\%)
 & 269(71.2\%)
\\
Total bursts 
 & 433
 & 399
 & 378
\\
\bottomrule
\end{tabular}
\end{center}
\footnotesize{\quad FIA represents the first incentive activity in the bursts.}

\end{table}

\section{Related Work} 
Power law distributions have been observed in various domains and contexts, such as biology \cite{saravia2018power}, general science \cite{martin2021implicit,meir2023efficient}, economics \cite{etro2018power,fricke2015distribution} and the social sciences \cite{barabasi2013network}. Many human behaviors, such as the intervals between sending emails \cite{malmgren2008poissonian} and the pattern of phone calls \cite{jiang2013calling}, have also been identified as following power-law distributions. Our work has discovered that the participation frequency of the population and the intervals between individual participation in activities exhibit power-law distributions in the context of community sports organizations.

Over the years, there have been continuous efforts to propose diverse models aimed at replicating and explaining data characterized by power-law distributions. Barab{\'a}si proposed the classic preferential attachment model, which can generate data exhibiting a power-law distribution with an exponent of 3 \cite{barabasi2013network}. There are also derivative models that can generate data with power-law distributions with exponents between 2 and 3 \cite{albert2000topology}. They have been widely used to explain the power-law distribution of node degrees observed in social networks. The decision-based queuing process \cite{barabasi2005origin} simulates the power-law distribution of waiting times for emails by randomly assigning priorities to each incoming task and following a rule of processing tasks in priority order. This suggests that the power-law distribution of waiting times for emails may be attributed to human decision-making based on priorities. The preferential attachment model suggests that the power-law distribution of node degrees in networks may be due to the preferential connection of newly added nodes to high-degree nodes in the network \cite{barabasi1999emergence}. In our HFBI model, the habit formation mechanism exhibits similarities to the preferential attachment model and can be proven to generate data conforming to a power-law distribution. In addition, the behavioral inertia component of the HFBI model introduces effective modifications, leading to a slight decrease in the exponent of the data while preserving its essential power-law characteristics.

Community sports organizations have been receiving increasing attention for their significant contributions to public health and social harmony. Klenk \emph{et al}. \cite{klenk2019social} investigated the participation of people with disabilities in community sports activities from three aspects: (1) social contacts, interactions, and friendships, (2) self-perception and identity formation, and (3) social acceptance, support, and embeddedness. Hanlon \emph{et al}. \cite{hanlon2022building} conducted a questionnaire survey to investigate the needs and initiatives for women's participation in community sports activities. Zhou \emph{et al}'s survey \cite{zhou2020community} revealed a correlation between the provision of community-sport services (both core and peripheral services) and participants’ satisfaction levels. To the best of our knowledge, there is no research that explores and comprehensively understands individual participation in community sports organizations from a data-driven and modeling approach.

\section{Conclusion}
Our study has identified new members of the power-law data family, a) the frequency of community sports participation among populations, and b) the interval of individual activity participation. The participation frequency exhibits a power-law distribution with a tail cutoff and an exponent less than 2. We have proposed HFBI - a model based on habit formation and behavioral inertia, to uncover the underlying causes for this power-law distribution. In the model, the behavioral inertia mechanism effectively complements the habit formation mechanism, with which alone one can only generate power-law distributions with an exponent greater than 2. The model provides a robust fit to the empirical data. Furthermore, Individual participation in community sports activities exhibits a burst-quiet pattern. Importantly, our study suggests that periods of high activity bursts are often driven by incentive activities, highlighting the importance of incentive activities to sustain long-term physical activity behavior.

Our results have important implications for the design of interventions aimed at promoting sustainable physical activity behavior. Interventions can be better tailored to align with individuals' behavioral tendencies by gaining insights into habit formation, behavior inertia, and incentive activities. Additionally, the classic preferential attachment process restricts the power law exponent to $\gamma>2$ \cite{ghoshal2013uncovering}, while many real-world networks exhibit $\gamma<2$ \cite{zhang2015exactly}. Our HFBI model based on habit formation and behavior inertia can be valuable in other domains where power-law distributions with low power-law exponents are observed, such as the population of cities \cite{rozenfeld2011area}, short-message communication \cite{wei2009heavy}, and corporate innovative patent counts \cite{choi2020power}. 

Despite the strengths of our study, there are limitations that should be noted. First, our study only focused on a sports community in a university, whose members are mostly well-educated university faculties and staff members, and may differ in the perception of self-motivated exercise from the population in the society at large. Further research is needed to understand how our study may be generalized to other community sports organizations. Secondly, the model cannot capture the behavior of extremely rare individuals who engage in activities excessively. As reported in the study's 80/20 rule, active individuals make a significant contribution to community activity participation, and future research should pay more attention to this group.

In conclusion, our study provides novel insights into the principle underlying human participation in community sports activities and offers practical implications for the design of interventions to promote sustained physical activity behavior and human health. Our findings may also have broader implications for other fields where power-law distributions are commonly observed.

\section*{Acknowledgments}
We would like to thank every member of the SJTU Health Community for their selfless commitment in building a supportive community and providing help to those in need.

\bibliographystyle{IEEEtran}
\bibliography{reference}

\end{document}